\documentclass[%
superscriptaddress,
amsmath,amssymb,
aps,
prfluids,
preprint,
]{revtex4-2}
\usepackage{CJK}
\usepackage{float}
\usepackage{overpic}
\usepackage{epstopdf}
\usepackage{graphicx}
\usepackage{dcolumn}
\usepackage{bm}
\usepackage{epsfig,dsfont,amssymb,amsmath,amsthm,amsfonts,amsbsy,mathrsfs}
\usepackage{siunitx}
\usepackage{url}

\usepackage{hyperref} 
\hypersetup{hypertex=true,
	colorlinks=true,
	linkcolor=blue,
	anchorcolor=blue,
	citecolor=blue}

\begin{document}

\begin{CJK*}{UTF8}{gbsn}
\title{Localized jammed clusters persist in shear-thickening suspension subjected to swirling excitation}

\author{Li-Xin Shi ({\CJKfamily{gbsn}石理新})}


\author{Song-Chuan Zhao ({\CJKfamily{gbsn}赵松川})}
\email[]{songchuan.zhao@outlook.com}
\affiliation{State Key Laboratory for Strength and Vibration of Mechanical Structures,\\ School of Aerospace Engineering, Xi'an Jiaotong University, Xi'an 710049, China}



\begin{abstract}

We investigate the dynamic evolution of heterogeneity in shear-thickening suspensions subjected to swirling excitation with a free surface. The uniform state of such a system may lose its stability when the oscillation frequency is above a threshold, and density waves spontaneously form (Shi \textit{et al.} JFM 2024). Here, we report a novel state where jammed clusters emerge in high-density region of the density waves. The jammed cluster exhibits unique motion, creating downstream high-density regions distinct from previously reported state of density waves. Additionally, theoretical calculations show that reducing suspension thickness lowers the frequency and global concentration $\Phi$ threshold for the heterogeneity onset. Notably, the minimal $\Phi$ for instability can be lower than the onset of discontinuous shear thickening transition. We also highlight the role of the free surface in cluster growth and persistence.


\end{abstract}


\maketitle
\end{CJK*}

\section{Introduction}
\label{sec:headings} 

Dense suspensions of micron-sized solid particles in a viscous liquid can exhibit a diverse array of rheological behaviors when subjected to applied shear stress~\cite{guazzelli2018rheology,morris2020shear}, such as shear thinning, shear thickening, and shear jamming. Among these behaviors, shear-thickening remains a subject of considerable debate regarding physical modeling and understanding its underlying mechanisms. One proposed cause of shear thickening is the formation of hydroclusters ~\cite{wagner2009shear,gurnon2015microstructure}. However, recent research indicates that the physical contact between particles likely plays a significant role in this phenomenon~\cite{singh2020shear,wyart2014discontinuous,seto2013discontinuous}.
As a result, the dominant interaction between particles shifts from hydrodynamic lubrication to frictional forces when the applied stress surpasses the inter-particle repulsion~\cite{morris2020shear}. 
Recently, a phenomenological model based on a mean-field perspective was introduced by Wyart and Cates~\cite{wyart2014discontinuous}.  Although this approach has yielded a range of results that have proven effective in describing both simulation and experimental data~\cite{singh2018constitutive,guy2018constraint,clavaud2017revealing,guy2015towards,comtet2017pairwise}, many questions remain. 

One set of those questions includes spatial as well as temporal dynamics within the shear thickening regime. Temporal fluctuations in bulk viscosity have been observed in classical rheology,  and visual evidence suggests the presence of associated spatial heterogeneities. These heterogeneities include transient bands that propagate along the vorticity direction~\cite{saint2018uncovering,rathee2022structure}, periodic density waves moving in the flow direction~\cite{ovarlez2020density,shi2024emergence}, as well as localized high viscosity or fully jammed phases ~\cite{rathee2017localized,rathee2020localized}. Direct observations on the suspension surface within a rheometer during shear have revealed local dilation-induced surface deformations~\cite{maharjan2021relation}.  Similarly, numerical simulations have identified stress fluctuations occurring at both the particle level and larger scales~\cite{nabizadeh2022structure,goyal2022flow,van2023minimally,rahbari2021fluctuations}. Heterogeneity has conventionally been associated with the negative slope of the constitutive curve under rigid confinement ~\cite{chacko2018dynamic}. However, while the role of boundary confinement in the shear thickening transition is proposed~\cite{Brown_2014,brown2012role}, further research is needed to explore the origin of instability transitions under free surfaces and the influence of boundary conditions on the growth of heterogeneity.

In our recent study, a dense cornstarch suspension was subjected to swirling excitation using an orbital shaker at a given suspension thickness, and a phenomenon of self-organization of density waves into a hexagonal pattern was reported~\cite{shi2024emergence}. Our interpretation of the origin of this instability centered around the competition between kinematic instability and particle migrations. In the current investigation, within the high fraction region, a new state with the emergence of jammed clusters is observed. The jammed clusters are characterized by a constant neighboring relation among closely packed particles. As the oscillation frequency increases, these jammed clusters grow to larger ones, eventually replacing the density waves across the system. Notably, reducing the suspension thickness $h_0$ also decreases the onset frequency of the jammed cluster. Moreover, it is noteworthy that the long-standing jammed clusters undergo decomposition when subjected to confinement by a fixed acrylic plate.


\section {Experimental set up and phenomenon}

\begin{figure*}[h]
	\centering
	\includegraphics{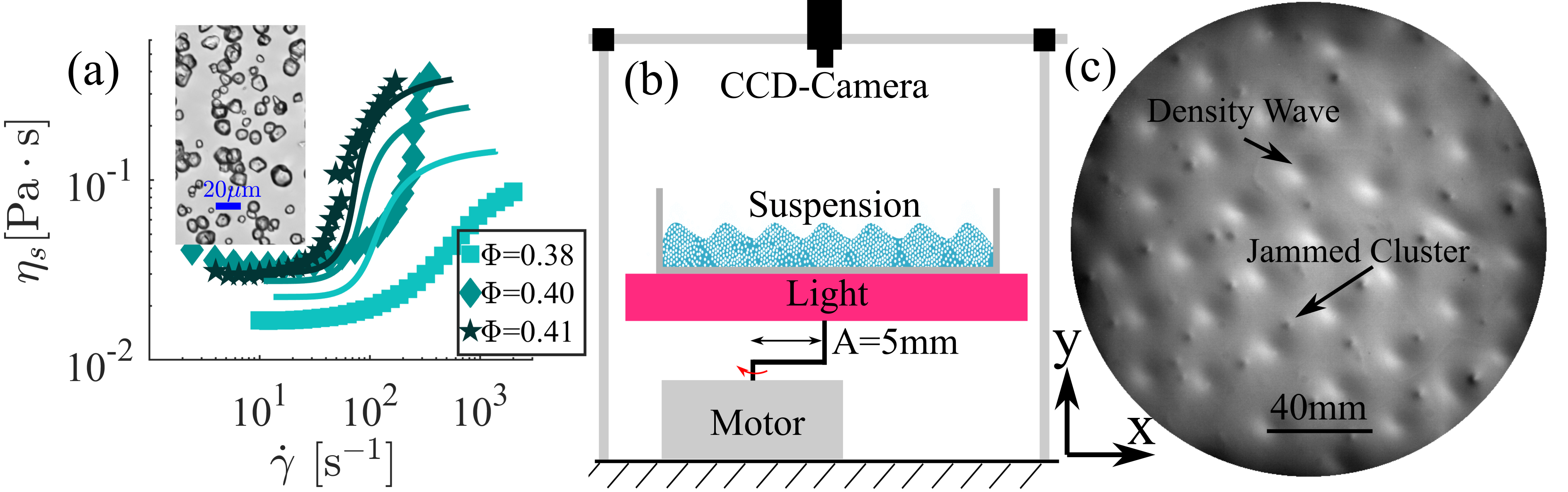}
	\caption{(a) Rheological data of the aqueous cornstarch suspension at different volume fractions. The solid curves are best fits of the Wyart-Cates model (see text). (b) Sketch of the set up. (c) Snapshot of the cornstarch suspension with $\Phi=0.43$ and $h_0=5~\si{mm}$ at $f=6~\si{Hz}$. Note that, jammed clusters arise in the high-$\phi$ region of density waves. Inset: Image of the cornstarch grains.}
	\label{fig:uniform_densitywave}
\end{figure*}

In our experimental setup, we utilize an open cylindrical container filled with a cornstarch suspension. The container has a total height of $50~\si{mm}$ and an internal radius of $160~\si{mm}$. The thickness of the suspension, $h_0$, is a variable in this study. The container is securely affixed to an orbital shaker (Heidolph Unimax 1010), which imparted a horizontal orbital motion-a circular movement of the entire platform. The oscillation frequency, denoted as $f$, can be adjusted within the range of $0.5$ to $8.33~\si{Hz}$, with a fixed amplitude of $A = 5~\si{mm}$, representing the radius of the orbital motion, as depicted in Fig.~\ref{fig:uniform_densitywave}b. This oscillation can be described as the superposition of periodic motion in the $x$ and $y$ directions: $x=A\mathrm{sin}(\omega t)$ and $y=-A\mathrm{cos}(\omega t)$, where $\omega=2\pi f$. We neglect the influence of the side wall, as reducing the container diameter does not alter the phenomena studied in this work. The system is illuminated from beneath by a red LED panel. A high-speed camera (Microtron EoSens 1.1cxp2) is securely mounted in the laboratory frame of reference, capturing the dynamics of the suspension and the light transmission. In our experiments, the variation in local density is the primary cause of light intensity contrast. Readers may refer to Ref.~\citenum{shi2024emergence} for the calibration data and method justification.

The suspension consists of cornstarch particles suspended in deionized water, with the addition of CsCl to achieve density matching. The cornstarch particles exhibit irregular shapes and have diameters ranging from 5 to 20 $\ \si{\mu m}$, as depicted in the inset of Fig.~\ref{fig:uniform_densitywave}a. The mass density of dry starch particles, denoted as $\rho_p$, is $1.61~ \si{kg/m^3}$. The rheological properties of the suspension are characterized using a stress-controlled Couette rheometer (Anton Paar 302) for different values of $\Phi$. The resulting flow curves are represented as $\eta(\tau)={\tau}/{\dot{\gamma}}$. The flow curves are fitted using the Wyart-Cates model: $\eta_{s}(\Phi)=\nu_0 \rho_s (\phi_J(\tau)-\Phi)^{-2}$~\cite{wyart2014discontinuous}. This mean-field model suggests that the jamming fraction $\phi_J$ undergoes a transition from its frictionless value $\phi_0$ to $\phi_m$ for frictional contacts when the applied stress $\tau$ exceeds a characteristic value $\tau^{*}$,  \textit{i.e.,}  $\phi_{J}=\phi_{0}-e^{{-\tau^{*}}/{\tau}}(\phi_0-\phi_m)$. 
The key parameters, $\phi_{0}$, $\phi_m$, and $\tau^*$, are obtained by fitting the rheological data. Although these parameters are assumed to be constant in the model, variations are observed when fitting data from different $\Phi$ independently~\cite{shi2024emergence}, implying a dependence of the parameters on $\phi$. Revisions to the model have been suggested to address this discrepancy~\cite{Thomas2020,Royer2016}. In this study, we follow the original Wyart-Cates model and fit all the rheological data collectively (see Fig.~\ref{fig:uniform_densitywave}a). The resulting values, $\nu_0=0.93\times 10^{-6} \si{m^2/s}$, $\tau^*=3.6\si{Pa}$, $\phi_{0}=0.45$, and $\phi_m=0.58$, will be used in subsequent analyses.
The Discontinuous Shear Thickening (DST) boundary $\tau(\Phi)$ must satisfy $(\phi_0-\Phi)/(\phi_0-\phi_m)=e^{-\tau^{*}/\tau} (1+2\tau^{*}/ \tau)$, leading to the minimal packing fraction for the onset of DST $\Phi_{\mathrm{DST}}=\phi_0 - 2e^{-1/2}(\phi_0-\phi_m)=0.43$ for the cornstarch suspension used in our experiments. The packing fraction $\Phi$ is varied within the range of $0.36$ to $0.44$ in experiments.

The uniform state of the suspension is unstable under bottom shear, and the development of inhomogeneity undergoes two distinct stages as the oscillation frequency increases. For instance, in a specific system characterized by $\Phi=0.43$ and $h_0=5~\si{mm}$,  the inhomogeneity first manifests as density waves when the oscillation frequency surpasses $2.33~\si{Hz}$. With a further increase in the oscillation frequency beyond $3.67~\si{Hz}$, numerous jammed clusters emerge in the high-$\phi$ region (Fig.~\ref{fig:uniform_densitywave}c). Notably, reducing  $h_0$ to $3.1~\si{mm}$ shifts the onset of both stages towards lower frequencies and $\Phi$. As jammed clusters signify the evolution of instability, how $h_0$ affects the instability of the uniform state is examined first.

\section {The effect of suspension thickness on the onset of heterogeneity}\label{Sec3}

Noting that the onset of instability in our suspension system exhibits characteristics akin to a compression wave of density, a two-fluid model is commonly employed to investigate its flow instability~\cite{chacko2018dynamic,Anderson1995,Batchelor1988}. For a comprehensive understanding of the model and linear stability analysis of this system, readers may refer to our previous study~\cite{shi2024emergence}. In this section, we primarily provide an intuitive argument for the stability criterion and present key findings related to analyzing the influence of $h_0$ for the stability criterion. We first consider the uniform state with respect to which the instability develops.

\begin{figure*}[h]
	\centering
	\includegraphics[scale=1]{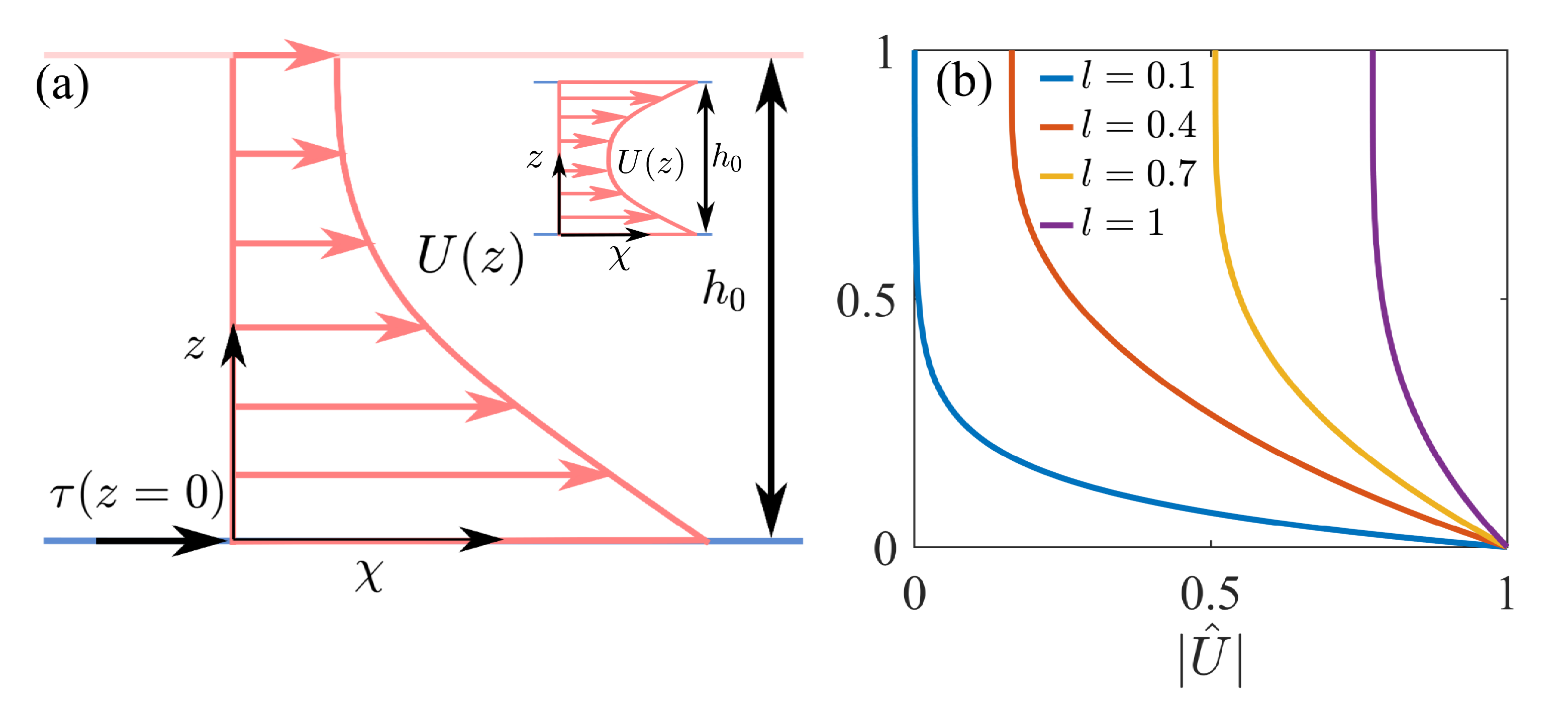}
	\caption{(a) Sketch of the notations, where $\chi$ represent the main flow direction. (b) Velocity profiles of the primary flow in  Eq.~\ref{eq.solution_of_stokes}. Inset: Velocity profile under the co-moving configuration (detailed in section \ref{Sec4}), where the suspension is covered by an acrylic plate that moves synchronously with the container.}
	\label{fig:velocity_profile}
\end{figure*}

Under the swirling excitation considered here, the flow velocity, $U(z,t)=U_x+iU_y$, of the suspension corresponds to a solution to the Stokes problem in two dimensions of uniform state under bottom circular shear, as expressed by:
\begin{equation}
	\hat{U}(\hat{z},\hat{t}) =e^{\frac{-\hat{z}}{l}}\left(\frac{e^{\frac{2(1+i)}{l}}+e^{\frac{2\hat{z}(1+i)}{l}}}{1+e^{\frac{2(1+i)}{l}}}\right)e^{i(\hat{t}-\frac{\hat{z}}{l})}.
	\label{eq.solution_of_stokes}
\end{equation}
Here, $l=\sqrt{2\nu/\omega}/h_0$, $\hat{U}=U/(A\omega)$, $\hat{z}=z/h_0$, $\hat{t}=t\omega$, where $\nu$ is the effective kinematic viscosity of the suspension. The terms $e^{-\hat{z}/l}$ and $e^{i(\hat{t}-\hat{z}/l)}$ represent the amplitude decay and phase lag, respectively. Notably, $l \sim \nu^{\frac{1}{2}}$ is prominently featured in the terms involving $\hat{z}/l$. Therefore, at high viscosity, the motion of the bottom plate deeply percolates to the surface of the suspension with minimal phase lag, as illustrated in Fig.~\ref{fig:velocity_profile}b.

The shear stress magnitude,
\begin{equation}
	\tau=\rho\nu\vert\partial U/\partial z\vert=\frac{l}{\sqrt{2}}\sqrt{\frac{\si{cosh}(2(\hat{z}-1)/l)-\si{cos}(2(\hat{z}-1)/l)}{\si{cosh}(2/l)+\si{cos}(2/l)}}\rho A h \omega^{2},
	\label{eq.shear_stress}
\end{equation}
is defined for a given $\Phi$. The viscosity of the suspension thus can be calculated using the constitutive relation, for which we employ the Wyart-Cates expression~\cite{wyart2014discontinuous}: $\nu(\Phi,\tau)=\nu_0 [\phi_0 (1-e^{-\tau^{*}/\tau})+\phi_m e^{-\tau^{*}/\tau}-\Phi]^{-2}$, with $\nu_0$, $\tau^{*}$, $\phi_0$ and $\phi_m$ obtained from the independent rheological measurement. Note that $\tau$ and $\nu$ are mutually dependent for given $\omega$ and $\Phi$. Additionally, the nonlinear velocity profile leads to a stress gradient along the vertical direction, potentially promoting the aggregation of particles near the surface. For simplicity, we neglect the secondary flow in the vertical direction and assume $\Phi$ is maintained across the entire system. To characterize the mainstream flow of the suspension, we employ $\tilde{U}(l,\Phi) = \vert U(z=h,t)\vert$ and $\Pi(l,\Phi) = \tau(z=0)$. The main flow is simplified to a one-dimensional homogeneous flow, which is closed by the constitutive relation $\nu(\Phi,\Pi)$. 

Once a disturbance of local density, $\phi$, arises within the mainstream flow, the resultant accumulation rate of particles is characterized by relative velocity between the loose and dense regions, $\Phi\tilde{U}^{\prime}$, where $\tilde{U}^{\prime}=\mathrm{d}\tilde{U}/\mathrm{d}\phi\vert_{\phi=\Phi}$. On the other hand, the pressure gradient $\nabla\Pi=\Pi^{\prime}\nabla\phi$ drives the particles to migrate out of the high $\phi$ regions. The corresponding migration velocity is characterized by $\sqrt{\Pi^{\prime}/\rho_p}$. When the accumulation rate of particles exceeds their migration, the inhomogeneous distribution of $\phi$ grows. A dimensionless threshold, ${\mathcal{C}}$, of order unity then sets the onset of instability of the uniform state:
\begin{equation}
	\tilde{\mathcal{C}} = \frac{\Phi  \tilde{U}^{\prime}}{\sqrt{\Pi^{\prime}/\rho_p}}>{\mathcal{C}}\sim\mathcal{O}(1).
	\label{eq.instability}
\end{equation}
The threshold ${\mathcal{C}}\sim\mathcal{O}(1)$ is to be determined by comparisons with experimental data and accounts for the features simplified in the model. 


\begin{figure*}[h]
	\centering
	\includegraphics[scale=1]{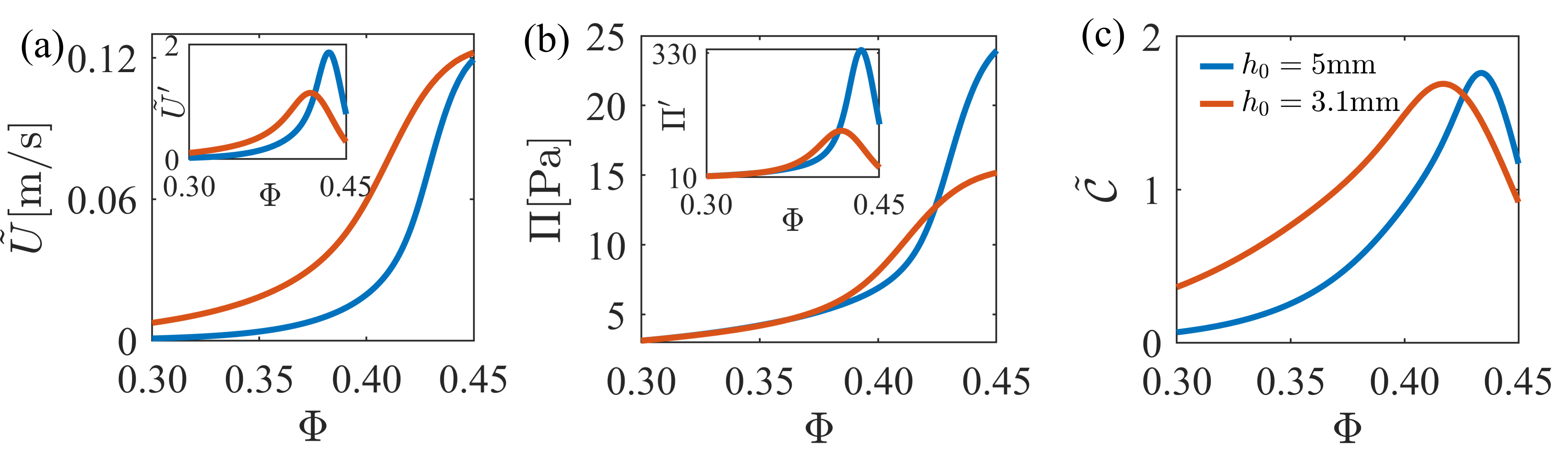}
	\caption{(a) The surface velocity $\tilde{U}$ increase with $\Phi$ at a fixed angular frequency $\omega=25~\si{rad~s^{-1}}$. (b) Shear stress at the bottom $\Pi$ increase with $\Phi$ at the same angular frequency. (c) The left-hand side of Eq.~\ref{eq.instability}, $\tilde{\mathcal{C}}$, varies with $\Phi$ at the same angular frequency. When $\tilde{\mathcal{C}}$ exceeds 1, the system theoretically becomes unstable. The minimum packing fraction at which $\tilde{\mathcal{C}}$ first exceeds 1 is reduced by decreasing $h_0$.}
	\label{fig:V_P_phi}
\end{figure*}

\begin{figure*}[h]
	\centering
	\includegraphics[scale=1]{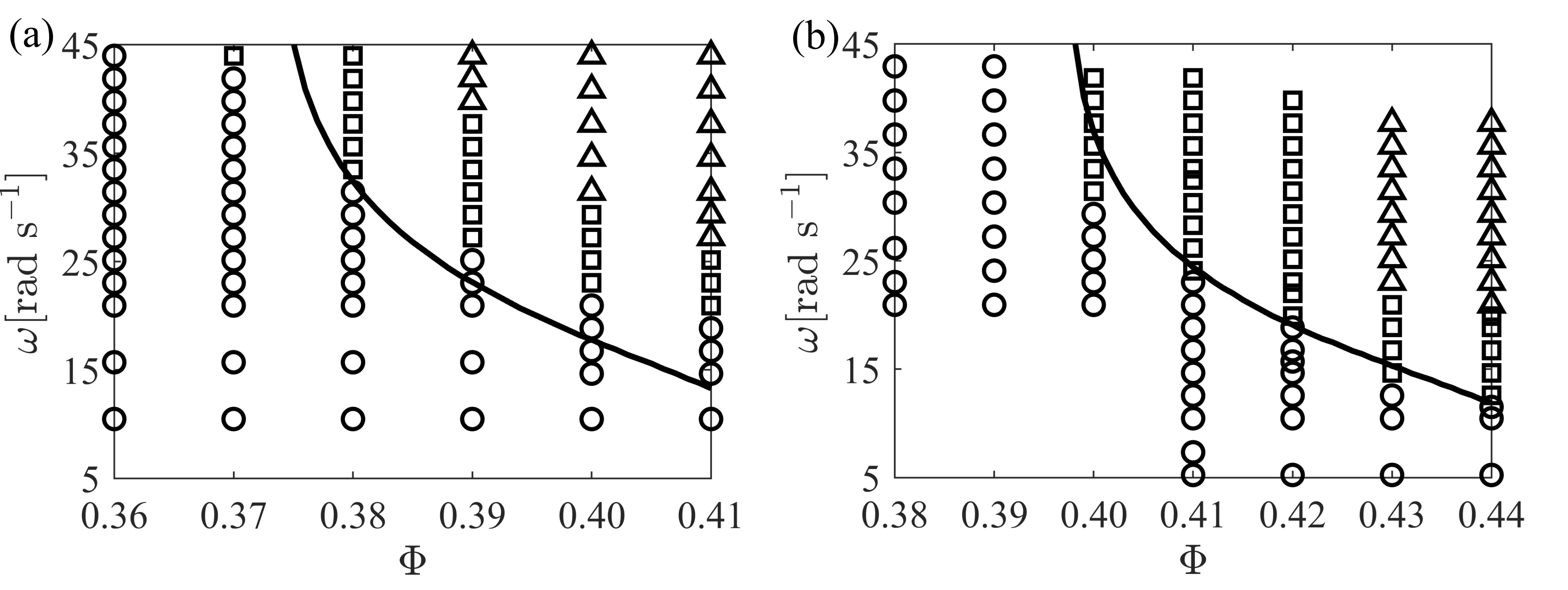}
	\caption{State diagram. Circles: uniform state. Squares: density waves. Triangles: jammed clusters. Black curve: theoretical onset frequency of instability growth. (a) $h_0 =3.1~\si{mm}$ and $\mathcal{C}=1.35$. (b) $h_0 =\si{5~mm}$ and $\mathcal{C}=1.15$.}
	\label{fig:State_diagram}
\end{figure*}

It is clear that the instability onset primarily hinges on the terms ${\tilde{U}}^{\prime}$ and ${\Pi}^{\prime}$ in Eq.~\ref{eq.instability}.
For a fixed angular frequency, the solution for $\tilde{U}$ in Eq.~\ref{eq.solution_of_stokes} exhibits a rapid increase at intermediate $\Phi$, corresponding to a regime with a large $\tilde{U}^{\prime}$ (Fig.~\ref{fig:V_P_phi}a).  An increase of $l$ would shift this regime toward lower packing fractions. The pressure term $\Pi$ follows a similar trend (Fig.~\ref{fig:V_P_phi}b). However, $\tilde{U}^{\prime}$ increases more rapidly than $\Pi ^{\prime}$, leading to $\tilde{\mathcal{C}}$, the left-hand side of Eq.~\ref{eq.instability}, increases with $\Phi$ at the uniform sate (Fig.~\ref{fig:V_P_phi}c). A reduction in $h_0$, leading to an increase in the parameter $l$, shifts the instability onset towards lower values of $\Phi$, as illustrated in Fig.\ref{fig:V_P_phi}c. Consequently, the boundary line representing the onset frequency of instability, $\omega_c (\Phi)$, moves to the lower $\Phi$ when $h_0$ is reduced from $5$ to $3.1~\si{mm}$ (Fig.~\ref{fig:State_diagram}). This shift results in the minimum packing density for persistent heterogeneity, $\Phi_{\mathrm{min}}=0.37$, which is notably below $\Phi_{\mathrm{DST}}=0.423$.
Previous studies suggest that the instability transition of shear-thickening suspensions under rigid confinement is closely linked to the characteristic $d\tau/d\dot{\gamma}\leq~0$ in the flow curve~\cite{chacko2018dynamic,rathee2022structure,saint2018uncovering}. However, within the scope of our investigation with free surface, instability can also be observed at a packing fraction that does not fall within the DST regime. Thus, a negative slope in the constituent curve is not a strict requirement for the initiation of heterogeneity. 

Beyond the stability criterion of the uniform state, the heterogeneity develops first into density waves and then into jammed clusters as the oscillation frequency increases. The influence of $h_0$ on the onset of these two states follows the same trend in the parameter space, as depicted in Fig.~\ref{fig:State_diagram}. Since the jammed clusters represent the further development of the heterogeneity, this state displays characteristics that cannot be captured by the analysis in the proximity of the uniform state. We investigate its dynamic features next.

\section{The long-standing jammed clusters}\label{Sec4}

\begin{figure*}
	\centering
	\includegraphics[scale=1]{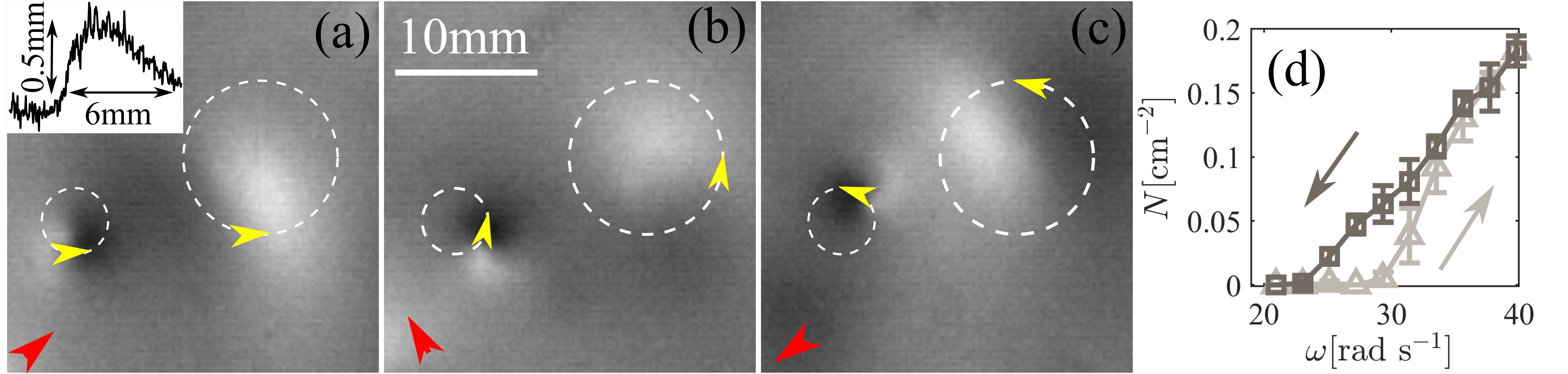}
	\caption{(a-c) A zoomed area at different phases in one oscillation period ($\Phi=0.43$, $h_0=5~\si{mm}$, $f=6~\si{Hz}$). The dashed
		circle denotes the trajectory of the center of the dense area, where the small circle indicates cuspidal point and the large one marks density wave. The yellow arrow indicates the instantaneous flow direction. The red arrow refers to the instantaneous drive direction. (d) The number of jammed clusters per unit area varies with oscillation frequency $\omega$. The upward and downward arrows correspond to the increase and decrease in oscillation frequency, respectively. The error bars are calculated by repeating the loop four times.}
	\label{fig:jammedcluster_densitywave}
\end{figure*}

Numerous cuspidal points emerge and persist in the high-$\phi$ region of the density waves as the oscillation frequency increases (dark/bright regions correspond to high/low $\phi$ areas~\cite{shi2024emergence}). These cuspidal points gradually grow and replace the density waves with the further increase of the oscillation frequency, eventually distributed throughout the system (Movie 1). The motion of individual cuspidal points follows circular paths and demonstrates the identical frequency of the external excitation, similar to density waves. Nevertheless, it exhibits distinct characteristics.
The circular motion of cuspidal points displays a phase lag of $\pi/8$ relative to density waves. Moreover, the typical diameter of the circular path of a cuspidal point measures $4.81~\si{mm}$, considerably smaller than the orbit of the container. In comparison, the typical diameter of the circular path of a density wave is $11~\si{mm}$. In addition, the steep density gradient is located upstream for the jammed cluster,  contrasting with the density wave. The peak-to-valley surface height difference is $0.5~\si{mm}$ (Fig.~\ref{fig:jammedcluster_densitywave}a, Inset), smaller than the $1~\si{mm}$ height difference associated with the density wave ~\cite{shi2024emergence}. The number of cuspidal points per unit area is recorded while ramping the oscillation frequency, and the hysteresis is consistently observed (Fig.~\ref{fig:jammedcluster_densitywave}d). This hysteresis can be attributed to the microscopic adhesion reported when cornstarch particles come into  prolonged contact~\cite{galvez2017dramatic}, suggesting the formation of permanent aggregates by interacting particles. Therefore, the cuspidal point is distinguished from the density wave state, a transient aggregate state that does not display hysteresis across its onset~\cite{shi2024emergence}.

\begin{figure*}[h]
	\centering
	\includegraphics[scale=1]{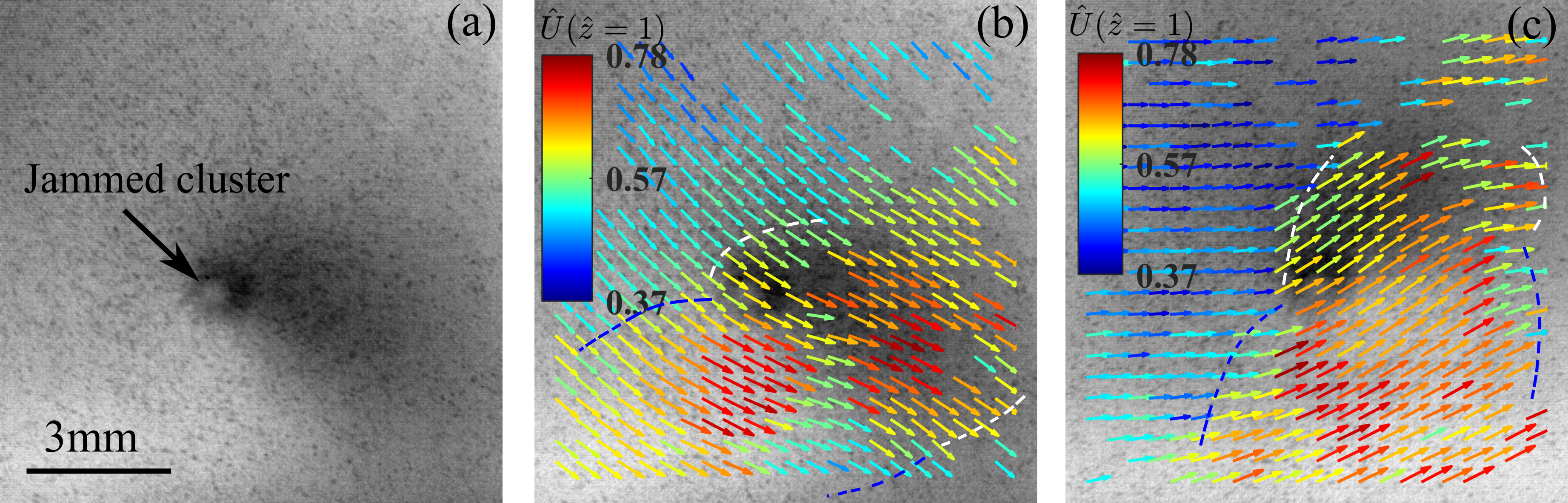}
	\caption{(a) A top-view snapshot of the jammed cluster. ($\Phi=0.41,~ h_0=2~\si{mm},~f=5~\si{Hz}$) (b-c) The scaled velocity of the surface $\hat{U}(\hat{z}=1)$  at two different phases. The region bounded by blue dotted lines: the low-lying area. The region bounded by white dotted lines: the area where suspension is propelled forward by the cluster.}
	\label{fig:micro_jammedcluster}
\end{figure*}

To investigate the microscopic nature of the cuspidal point, we capture microscopic images around the cuspidal point at a resolution of $28.2~\si{\mu m/pixel}$ and calculate the flow field using particle image velocimetry (PIV), as illustrated in Fig.~\ref{fig:micro_jammedcluster}b-c. The multimedia content showcases the consistent arrangement of closely packed tracer particles within the cuspidal point (Movie 2), and the flow field reveals that the velocity of the cuspidal point remains constant at different phases (Fig.~\ref{fig:micro_jammedcluster}b-c). These observations strongly suggest the formation of a jammed cluster. Equation.~\ref{eq.solution_of_stokes} elucidates that the scaled velocity at the free surface, $\hat{U}(\hat{z}=1,l)$, increases with the parameter $l$ while simultaneously experiencing a decrease in phase lag. In this context, the jammed cluster experiences a acceleration, indicating a local surge in $l$. However, it is essential to note that the velocity of the jammed cluster remains smaller than that of the drive plate, suggesting that the jammed cluster has not percolated to the bottom. Furthermore, the flow field becomes distorted around the cluster. The suspension downstream of the cluster is propelled forward by the cluster (bounded by white dashed lines), generating a low-lying, low-density area beside the jammed cluster (bounded by blue dashed lines). The stress gradient drives the particles to fill this low-density region from the side, resulting in a rapid flow in this area.

As a jammed cluster emerges, the aggregate particles push against the interface, forming a hump-like cuspidal point visible on the surface of the suspension (Fig.~\ref{fig:jammedcluster_densitywave}a, Inset). 
Extensive research has reported the essential role of boundary confinement in shear thickening phenomena~\cite{brown2012role,cates2005dilatancy,Fall2008,Brown_2014,boyer2011unifying}. When the expansion of the granular phase is impeded by boundary constraints during shearing, it leads to the generation of normal stresses as particles exert pressure against the boundaries. The open system studied in our experiments is bounded by the air-suspension interfacial tension, $\sigma$, and the curvature of the free surface, $\Delta h/(W/2)^{2}$. For the jammed cluster, $\Delta h=0.5~\si{mm}$ represents the height difference, and $W=6~\si{mm}$ denotes the width of the jammed cluster (Fig.~\ref{fig:jammedcluster_densitywave}a, Inset). Therefore, the confining  pressure is approximately $\sigma_{\mathrm{Norm}}\simeq \sigma \Delta h/(W/2)^{2}\simeq 4.04~\si{Pa}$,  exceeding the repulsive force between particles $\tau^{*}=3.6~\si{Pa}$. This further substantiates that the particles within the jammed cluster can be in physical contact.


It is essential to emphasize that long-lived inhomogeneities have previously been reported only near free surfaces~\cite{ovarlez2020density,maharjan2021relation,gauthier2023shear}, indicating that boundary confinement influences the development of underlying clusters (however, when considering the onset of instability, we focus on a state close to uniformity, where the boundary confinement isn't crucial). We hypothesize that under soft confinement, where the boundary is sufficiently flexible to relax normal stress, clusters can grow larger and persist in the system, as the jammed clusters observed in our experiments. Conversely, rigid constraints completely suppress stress release, leading to a dramatic increase in the local stress when the clusters grow to a certain size, ultimately causing them to collapse under stress. To assess the influence of rigid confinement, a co-moving acrylic plate is introduced to cover the suspension, where the shear profile under the free surface is overlayed with its reflection (Fig.~\ref{fig:velocity_profile}a, inset). Under this configuration, the uniformity sustains even for $\omega>\omega_c$. However, upon removing the confinement and oscillating at $\omega<\omega_c$, a transient growth of clusters is observed~\cite{shi2024emergence}. Therefore, finite-size clusters exists (too small to be discerned by our method) under the rigid confinement, whose growth is, however, suppressed.

Recent experiments with spatial resolution have suggested that when the applied shear stress is beyond a critical value, $\tau_c$ ($\tau_c>\tau^{*}$), localized jammed regions emerge within the system, corresponding to shear-thickening transition. These jammed clusters rupture under shear, leading to transient high-stress events~\cite{rathee2017localized,rathee2020localized,rathee2022structure}. In our experiments, when the co-moving confinement is replaced by a fixed one, numerous high-density regions persist for less than one period (Movie 3). Under the stress-controlled scenario, the transient localized high-stress regions further influence the global shear rate, giving rise to complex fluctuations in its behavior~\cite{saint2018uncovering,ovarlez2020density}. From the perspective that normal stress is proportional to shear stress, there may exists a critical normal stress, $p_c$, corresponding to $\tau_c$, at which clusters could persist in the system as long as the interface constraints are maintained near that value.

\section{discussion and conclusion}

In this study, we investigate the dynamic evolution of heterogeneity in shear-thickening suspensions with a free surface under swirling excitation. As the suspensions lose their uniform states and become unstable, the heterogeneity grows. The stability criterion, denoted as ${\tilde{\mathcal{C}}}=\phi \tilde{U}^{\prime}/\sqrt{\Pi^{\prime}/\rho_p}$, is largely determined by the parameter $l=\sqrt{2\nu/\omega}/h_0$, the square of which is merely the inverse Reynolds number representing the significance of viscosity relative to inertia. The shear-thickening transition results in a substantial amplification of $ l$, effectively diminishing the impact of inertial effects. Reducing the suspension thickness $h_0$ has a similar effect on amplifying $l$. In consequence, the minimal packing fraction for instability ($\Phi_{\mathrm{min}}$) at small $h_0$ could become lower than the minimal packing fraction for Discontinuous Shear Thickening (DST). The presence of the $\mathsf{S}$-shaped flow curve is thus not an essential prerequisite for the onset of heterogeneity under free surface conditions. However, it may play a role in more developed states, such as jammed clusters, where the local density is dramatically increased.

The heterogeneity manifests as density waves just above the instability onset. Jammed clusters are formed far beyond the onset (Fig.~\ref{fig:State_diagram}). The upward migration of particles promotes their formation (which is neglected in section~\ref{Sec3} for analyzing the uniform state). The jammed cluster, moving as a collective entity, exhibits properties distinct from the density wave. The jammed clusters propel the downstream suspension to accelerate forward while creating a significant density gradient upstream, which is in stark contrast to that of density waves. During the process of reducing oscillation frequency, the disappearance of clusters noticeably lags behind their onset frequency, a typical hysteresis feature. In addition, in the experimental setup studied here, jammed clusters have been also observed in an aqueous solution of polydisperse silica beads (average diameter of $20~\si{\mu m}$), a purely repulsive shear-thickening system. Importantly, when electrolytes are dissolved within the solvent, which reduces the range of the repulsive forces below that of the particle roughness~\cite{clavaud2017revealing}, macroscopic heterogeneity is significantly weakened (see Appendix~\ref{appA}). Therefore, in the context of our study, the observed heterogeneity is a common phenomenon associated with shear thickening.

Finally, we delve into the importance of flexible confinement for the development of long-standing jammed clusters. The results from simulations ~\cite{goyal2022flow,van2023minimally,nabizadeh2022structure} and experiments~\cite{cheng2011imaging} reveal that finite-size clusters already form at the onset of shear-thickening transition under rigid confinement. These clusters suddenly grow to a size comparable to the gap between the boundaries and break as the stress increases beyond a critical value~\cite{Rathee2021,saint2018uncovering,rathee2017localized,rathee2020localized,rathee2022structure}, leading to a transient dramatic fluctuation in the local stress. However, in the case of soft confinement, such as the free surface studied in our work, the normal stress generated during shearing is effectively released, setting a cutoff on $\Pi'$ in the criterion (Eq.~\ref{eq.instability}). Therefore, it allows the long-standing clusters grow to larger ones driven by the dynamic instability. 
It is plausible that the relative significance between the shear driving and the boundary deformation further determines how instability manifests, either as density waves or jammed clusters. Further investigation into the role of boundary flexibility is needed to provide deeper insights into shear-thickening phenomena.

\section{acknowledgments}
This work is partially supported by the National Natural Science Foundation of China (Grant No. 12172277).

\appendix
\section{Polydispersion silica bead experiments}\label{appA}

\begin{figure}
	\centering
	\includegraphics[scale=1]{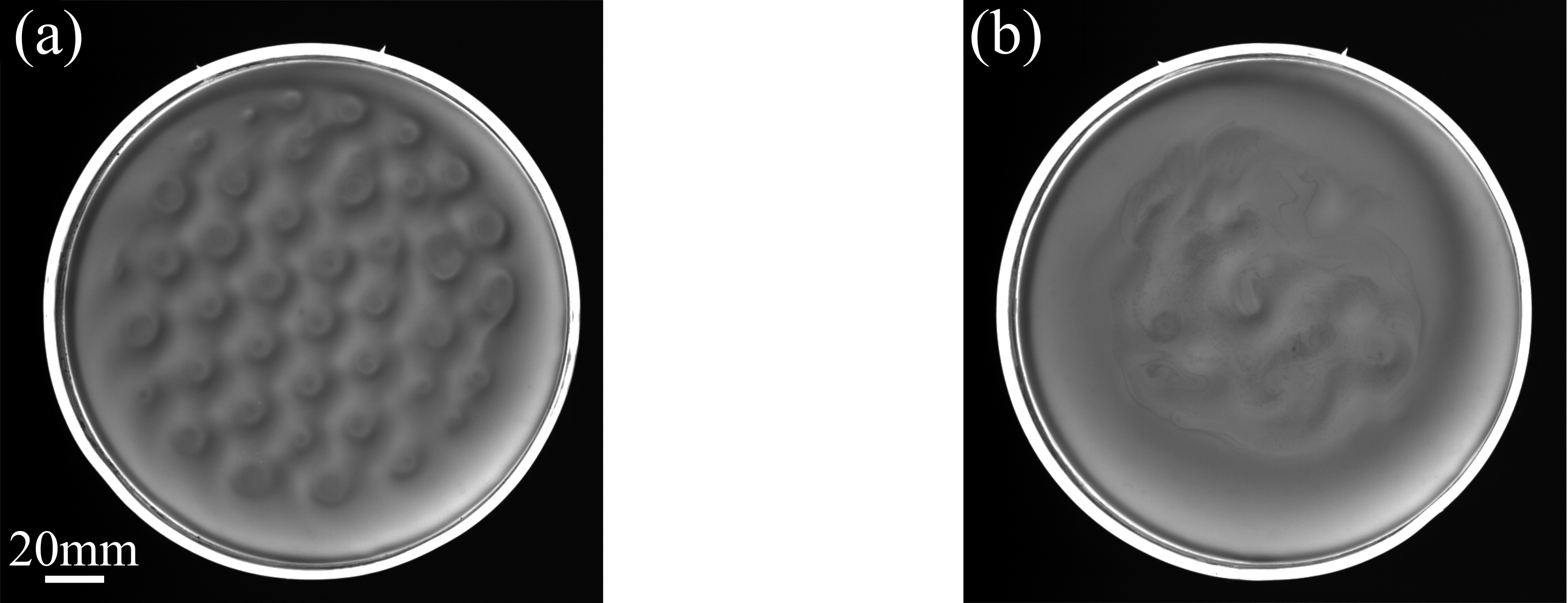}
	\caption{Snapshots of silica bead aqueous suspension under oscillation. ($\Phi=0.553$, $h_0=3.1~\si{mm}$, $f=8~\si{Hz}$) (a) deionized water solution without salt[NaCl], (b) $0.1~\si{m/L^{-1}}$ salt[NaCl] is added in the solution.}
	\label{fig:par_vel}
\end{figure}

We conduct alternative experiments using a polydisperse silica bead solution with an average grain diameter of $d_g= 20~\mu \si{m}$. As illustrated in Fig.~\ref{fig:par_vel}a, for a solution of deionized water,  jammed clusters are distributed throughout in the container. However, when the suspension is added with $0.1~\si{m/L^{-1}}$ salt[NaCl], the jammed clusters are noticeably suppressed (Fig.~\ref{fig:par_vel}b).

\bibliography{main}
\end{document}